# Colossal magnetocapacitance and scale-invariant dielectric response in phase-separated manganites

Ryan P. Rairigh, Guneeta Singh-Bhalla, Sefaatin Tongay, Tara Dhakal, Amlan Biswas and Arthur F. Hebard

*Department of Physics, University of Florida, Gainesville, FL 32611-8440*

A full characterization of phase separation and the competition between phases is necessary for a comprehensive understanding of strongly-correlated electron materials (SCEM), such as under-doped high temperature superconductors[1,2], complex oxide heterojunctions[3], antiferromagnetic[4] and ferromagnetic[5] spinels, multiferroics[6,7], rare-earth ferroelectric manganites[8], and mixed-valence manganites where phase competition is the dominant underlying mechanism governing the insulator-metal (IM) transition and the associated colossal magnetoresistance (CMR) effect[9-11]. Thin films of SCEMs are often grown epitaxially on planar substrates and typically have anisotropic properties that are usually not captured by edge-mounted four-terminal electrical measurements, which are primarily sensitive to in-plane conduction paths. Accordingly, the correlated interactions in the out-of-plane (perpendicular) direction cannot be measured but only inferred. We address this shortcoming and show here an experimental technique in which the SCEM under study, in our case a 600 Å-thick $(La_{1-y}Pr_y)_{0.67}Ca_{0.33}MnO_3$ (LPCMO) film, serves as the base electrode in a metal-insulator-metal (MIM) trilayer capacitor structure. This unconventional arrangement allows for *simultaneous* determination of colossal magnetoresistance (CMR) associated with dc transport parallel to the film substrate and colossal magnetocapacitance (CMC) associated with ac transport in the perpendicular direction. We distinguish two distinct strain-related direction-dependent insulator-metal (IM) transitions and use Cole-Cole plots to establish a heretofore unobserved collapse of the dielectric response onto a universal scale-invariant power-law dependence over a large range of frequency, temperature and magnetic field. The resulting phase diagram defines an extended region where the competing interaction of the coexisting ferromagnetic metal (FMM) and charge-ordered insulator (COI) phases[10-15] has the same behavior over a wide range of temporal and spatial scales. At low frequency, corresponding to long length scales, the volume of the phase diagram collapses to a point defining the zero-field IM percolation transition in the *perpendicular* direction.

The simultaneous measurement of the zero field ($H = 0$) parallel resistance $R_\parallel(T)$ and perpendicular capacitance $C(T)$ transport in the LPCMO film, which is embedded as the base electrode in the trilayer configuration shown schematically in Fig. 1a, is captured in the temperature-dependent curves of Fig. 1b. The two-terminal $C(T)$ measurements correspond to the real part $C'(\omega)$ of a complex lossy capacitance, $C^*(\omega) = C'(\omega) - iC''(\omega)$, measured at $f = \omega/2\pi = 0.5$ kHz. As temperature $T$ decreases from 300K, the prevailing high temperature paramagnetic insulator (PI) phase gives way near $T = 220$ K[14] to a dominant COI phase having a resistance that rapidly increases as $T$ continues to decrease. Minority phase FMM domains appear and begin to short circuit the resistance rise as $T$ continues to decrease. At the resistance peaks the percolative IM transition for transport in the parallel direction through FMM domains occurs at temperatures $T^\downarrow_{IM,\parallel} = 95$ K and $T^\uparrow_{IM,\parallel} = 106$K where the down($\downarrow$)/up($\uparrow$) arrows indicate the cooling/warming direction of the temperature sweep. Below $T^\downarrow_{IM,\parallel}$ the FMM phase rapidly dominates with decreasing $T$, and $R_\parallel$ decreases by four orders of magnitude. The reverse takes place upon warming through $T^\uparrow_{IM,\parallel}$

The $C(T)^{\downarrow\uparrow}$ traces reach plateaus at high and low temperature where the LPCMO is in its respective PI and FMM states, both of which have sufficiently low resistivity to act as metallic electrodes in a MIM structure. Between the $C_{AlOx}(T)$ plateaus, the $C(T)^{\downarrow\uparrow}$ traces appear to be roughly inverted replicas of $R_\parallel(T)^{\downarrow\uparrow}$ except for an ~20K shift to lower temperatures of the capacitance minima below the corresponding resistance maxima. At perpendicular fields $H = 50$kOe the capacitance (blue curve) has increased by a factor of 1000 above the zero-field minimum; colossal magnetocapacitance (CMC) is clearly present. The remnant capacitance dip at 50 kOe disappears at 70 kOe and the linear temperature dependence $C(T) = C_{AlOx}(T)$ is identical to that of separately measured Al/AlOx/Al capacitors. Deviations of $C(T)^{\downarrow\uparrow}$ below $C_{AlOx}(T)$ thus reflect the competition of FMM and COI phases, and it is here where the CMC effects occur.

To understand how measurement of $C(T)$ is more than just a complicated way of measuring $R(T)$, we analyze in detail circuit equivalents (see Supplementary Information) of the measurement configuration. We start by modeling the measurement configuration as a resistance $R_s$ in series with the parallel combination of $C^*(\omega)$ and a dc resistance $R_0$. We then justify this model by establishing two necessary and sufficient conditions to assure that $R_s$ can be ignored, namely: (1) the leakage resistance $R_0$ is significantly higher than any of the impedance elements in our configuration, and (2) frequencies are chosen such that $R_s$ is the smallest relevant impedance. Under these conditions the measured quantity, $C^*(\omega)$, is determined by voltage drops that are perpendicular to the film interface and the corresponding in-plane equipotentials depicted schematically by the dashed horizontal lines in the circuit schematic of Fig. 1c.

We model $C^*(\omega)$ using a Maxwell-Wagner (MW) circuit equivalent[16] (see Supplementary Information) in which the manganite impedance, expressed as a parallel combination of a resistance $R_M$ and capacitance $C_M$, is connected in series with a leak-free capacitance $C_{AlOx}$ representing the Al/AlOx circuit element. The qualitative agreement shown in Fig. 1b between the capacitance data (red) and the MW model calculation (green) confirms the appropriateness of the MW model as has also been shown in related dielectric studies of transition metal oxides[17], spinels[4,5,18] and multiferroics[7] where the material under investigation is the 'insulator' (I) of a MIM structure rather than the base electrode as discussed here.

The MW analysis uses the measured $R_\parallel$ as an input and therefore enforces an alignment in temperature (Fig. 1b) between the measured resistance maxima and the calculated capacitance minima. Since the equipotentials of the capacitance measurement are parallel to the film surface, the misalignment of the measured capacitance minima (~20K in Fig. 1b), can best be explained by assuming that $R_M$ in the MW calculation should be the perpendicular resistance $R_\perp(T)$ rather than $R_\parallel(T)$. Hence the measured capacitance minima are in alignment with putative resistance maxima corresponding to percolative IM transitions in $R_\perp(T)$. The IM transition temperatures in the perpendicular direction, $T^\downarrow_{IM,\perp}$=77 K and $T^\uparrow_{IM,\perp}$=91 K, are therefore equal to the temperatures where the capacitance minima occur at noticeably lower values than the corresponding temperatures, $T^\downarrow_{IM,\parallel}$=95K and $T^\uparrow_{IM,\parallel}$=106K, for the $R_\parallel(T)$ maxima.

Strain effects explain the occurrence of anisotropy and two separate percolation transitions. The LPCMO films were grown on $NdGaO_3$, which is known to stabilize the pseudocubic structure of the FMM phase at low temperatures[19,20]. Hence, at the film-substrate interface a strain-stabilized FMM phase is formed at low temperatures. The effect of the substrate diminishes away from the interface[19] as shown schematically by the shading of Fig. 1c, where the LPCMO electrode is depicted as an infinite RC network comprising local resistances $r_\perp$ and $r_\parallel$ respectively perpendicular and parallel to the interface. These distributed resistor elements are locally equal to each other but both increase as a function of distance away from the interface[21]. In the measurement of $R_\parallel$ the strain-stabilized FMM region "shorts out" the higher resistance state occurring away from the interface and percolation occurs at a higher temperature $T^{\downarrow\uparrow}_{IM,\parallel}$ than the percolation transition temperature $T^{\downarrow\uparrow}_{IM,\perp}$ as measured by the capacitance minima. For thicker films, the two IM transitions should converge to a single value representing isotropic 3D percolation of bulk LPCMO with the temperature difference $\Delta T^{\downarrow\uparrow}_{IM} = T^{\downarrow\uparrow}_{IM,\parallel} - T^{\downarrow\uparrow}_{IM,\perp}$ approaching zero. The convergence toward bulk properties ($\Delta T^{\downarrow\uparrow}_{IM} \rightarrow 0$) was experimentally confirmed (see Supplementary Information) with transport measurements on three additional samples with LPCMO thicknesses ranging from 300Å to 900Å.

To fully characterize the intrinsic dielectric response of the LPCMO film we utilize Cole-Cole plots in which the dielectric loss $C''$ is presented as a function of the polarization $C'$ while an external parameter, usually frequency, is varied[16]. We show such a plot on logarithmic axes in Fig. 2a at T=65K (warming) for the magnetic fields indicated in the inset. As the frequency is swept from low (50Hz) to high (20kHz) at each field, $C''(\omega)$ passes from a region where $C_{AlOx}$ dominates, subsequently reaches a peak value at $\omega R_M C_M=1$ (vertical arrow), and then enters the high frequency region where the intrinsic response of the manganite dominates and the data collapse onto the same curve. The low-to-high frequency crossover from no-collapse of the data to collapse is magnified in the inset.

At higher temperatures the data collapse is more striking as shown in the Cole-Cole plot of Fig. 2b. Independently of whether the implicit variable is frequency f=$\omega/2\pi$, $T$ or $H$, the dielectric response collapses onto the same curve when the remaining two variables are fixed (see inset). As $\omega$ increases, or equivalently, as $T$ or $H$ decrease, $C''$ approaches zero and $C'$ approaches a constant $C_\infty$ representing the bare dielectric response. We find that when $C''$ is plotted against $C'-C_\infty$ on double logarithmic axes, all the data collapse onto a straight line (inset) described by the equation,

$$C''(\omega,T,H) = \Lambda\left[C'(\omega,T,H) - C_\infty\right]^\gamma \qquad . \qquad \text{(Eq. 1)}$$

The three fitting parameters have values $\Lambda=9.404 \pm 0.049$ pF$^{0.3}$, $\gamma=0.701 \pm 0.004$ and $C_\infty=0.135 \pm 0.003$ pF. We note that Eq. 1 is a generalization of Jonscher's "universal dielectric response", which requires $\gamma = 1$ and describes well the high frequency dielectric response of most dielectrics[22]. As shown below, the generalized power-law scaling that we have observed with respect to three independent variables ($\omega$, $T$, $H$) is associated with a percolation transition in which FMM and COI phases form clusters that compete on self-similar length scales.

The universal power-law scaling collapse (PLSC) of the data described by Eq. 1 applies to a region of fTH space with boundaries that can be accurately determined from Cole-Cole plots which use $T$ as the implicit variable over the entire temperature range. As an example of our procedure, we show in Fig. 3 a Cole-Cole plot on linear axes with temperature (cooling) as the implicit variable for $f = 500$ Hz and magnetic fields identified in the legend. The dielectric response is divided into two branches by a pronounced crossover from PLSC behavior denoted by the orange line described by Eq. 1 (see also Fig. 2b inset) in common with the lower part of the upper branch to no-collapse (NC) behavior (lower branch). As temperature decreases on the upper branch, the data follow a non power-law collapse (NPLC) which at well defined field-dependent upper critical temperatures, $T^{\downarrow}_{upper}(H)$, merge onto the orange line representing PLSC behavior. As $T$ continues to decrease in the PLSC region, there is a second critical temperature, $T^{\downarrow}_{lower}(H)$, marking the demarcation points between branches by an almost 180º change in the direction of the trajectory as indicated by the "turnaround" arrow in the inset. These field-dependent crossover points precisely define $T^{\downarrow}_{lower}(H) = T^{\downarrow}_{Cmin}(H)$ where $C'(H)$ is minimum and which by our MW analysis has been shown to be the same as $T^{\downarrow}_{IM,\perp}(H)$. A similar analysis holds for warming curves.

The accurate determination of field-dependent critical temperature boundaries at each frequency allows us to delineate triangular-shaped areas in TH space where PLSC is obeyed. Such a phase diagram is shown for cooling in Fig. 4. The two critical temperature boundaries, $T^{\downarrow}_{upper}(H)$ and $T^{\downarrow}_{lower}(H)$, are determined by the NPLC-to-PLSC (closed symbols) and the PLSC-to-NC transition (open symbols). For comparison we show (crosses) the IM boundary $T^{\downarrow}_{IM,\parallel}(H)$, determined from the peaks in $R^{\downarrow}_{\parallel}(H)$. The roughly parallel offset from the lower boundary $T^{\downarrow}_{IM,\perp}(H)$ of the PLSC phase implies two different percolation transitions, one for $R_\perp$ and the other for $R_\parallel$. Increasing the frequency pushes $T^{\downarrow}_{upper}(H)$, to higher temperatures and fields, thus increasing the volume of $fTH$ space where PLSC occurs. For any TH region bounded by an upper critical boundary determined at $f=f_0$, PLSC is obeyed within that region for all $f \geq f_0$. Therefore as $f_0$ decreases and the corresponding length scales being probed increase, the corresponding TH region, where PLSC is obeyed for all $f \geq f_0$, shrinks to a point defined by the temperature $T^{\downarrow}_{IM,\perp}=77$ K at $H=0$ (marked by X on Fig 4) where $R_{\perp}$ is maximum and percolation occurs. A similar phase diagram shifted ~20K to higher temperatures occurs for warming data.

In summary we have presented a new technique uniquely suited to probing both the longitudinal dc transport and perpendicular ac dielectric response of thin films comprising a mixture of competing metallic and insulating phases. We expect that the technique, which advantageously includes information about dynamics, will find wide application in studies of a variety of anisotropic thin-film systems including those where the presence of competing phases is under debate. Candidates for study include layered manganites[23], under-doped high-$T_c$ superconductors in which superconductivity emerges from an insulating temperature dependence[1], electron-doped cuprate $Pr_{2-x}Ce_xCuO_4$ (PCCO) *c*-axis films[24] with bulk anisotropic resistivity ratios ($\rho_c/\rho_{ab}$) reported[25] to be as high as a factor of ten thousand, and c-axis graphite where resistivity ratios are typically a factor of one thousand[26]. We have also demonstrated that our technique can be sensitive to strain at epitaxial interfaces and thus capable of determining through thickness dependence studies the critical thickness for a relaxed top interface. By incorporating an LPCMO film as the 'metal' (M) base electrode of an MIM structure and using an ultra low leakage dielectric (AlOx) for the insulating spacer, we prevent the metallic phase of the LPCMO from shorting out the measurement as it would if it were the middle layer (I) of a conventional dielectric configuration. Circuit analysis shows, and experiment demonstrates, that over a well defined frequency range the large Al/AlOx capacitor acts as a baseline reference that 'decloaks' or makes visible the much smaller capacitance of the series-connected LPCMO film. By distinguishing two strain-related direction-dependent IM transitions, which merge as the film thickness increases and transport becomes isotropic, we confirm the sensitivity of our capacitance and resistance measurements to perpendicular and longitudinal transport and at the same time establish phase diagrams in fTH space for $T > T^{\downarrow\uparrow}_{IM,\perp}(H)$ where a surprising and heretofore unobserved power-law scaling collapse (PLSC) is observed. The power-law dependence implies scale invariance from short length scales determined by the highest measurement frequencies to significantly longer length scales, where the low-frequency PLSC regions collapse to the points $T^{\downarrow\uparrow}_{IM,\perp}(H=0)$ defining the dc cooling/warming percolation transitions. The factor of 1000 change in capacitance is too large to be caused by changes on the surfaces of the LPCMO and must therefore be attributed to the competition of microscopic FMM/COI clusters intrinsic to the bulk manganite. The shortest length scale in the PLSC region must be somewhat smaller than the LPCMO film thickness (600 Å); otherwise, it does not make sense to have a resistance gradient in the perpendicular direction. As temperature is lowered through $T^{\downarrow\uparrow}_{IM,\perp}(H)$, the competition of phases on microscopic (~100 Å) length scales in the PLSC region crosses over to a competition on macroscopic (~1 μm) length scales in which the metallic fraction of area covered increases rapidly from $C_{min}/C_{AlOx}=10^{-3}$ to unity as large clusters of the FMM phase dominate[10,11]. A full understanding of these results will be a challenge to the contrasting theories of phase separation and competition in manganites based on intrinsic disorder[12], long range strain interactions[13], blocked metastable states[14] or thermally equilibrated electronically soft phases[15].

**Methods**

The 600 Å-thick $(La_{1-y}Pr_y)_{1-x}Ca_xMnO_3$ (LPCMO) thin films were grown at a stoichiometry $x = 0.33$ and $y = 0.5$. The value of $x$ was chosen to maximize the Curie temperature of the FMM phase. The value $y = 0.5$ was chosen because at this composition thin films of LPCMO show a large hysteresis in resistivity in the cooling and warming cycles, which is a signature of phase coexistence. For thin films with $y < 0.4$ the hysteresis becomes negligible while for $y > 0.6$ the high resistance at $T_{IM}$ requires extremely low leakage resistance of the AlOx dielectric. The films were grown using pulsed laser deposition (PLD) at a rate of 0.05 nm/s on $NdGaO_3$ (NGO) (110) substrates kept at 820°C in an oxygen atmosphere of 420 mTorr. Standard θ-2θ x-ray diffraction data show that the films are epitaxial and of a single chemical phase.

The 100 Å-thick AlOx dielectric was deposited at a rate of 2-3 Å/min via rf magnetron sputtering of an alumina target using separate *ex situ* procedures previously described[27]. A critical step to assure a leak free insulating AlOx layer is the preconditioning of the alumina target in a partial pressure of oxygen prior to sputter deposition in pure argon ambient. The 1000 Å-thick circular Al counterelectrode with radius 0.4 mm was thermally evaporated through a shadow mask in a separate chamber.

The capacitance measurements (Fig. 1, red) were made using an Andeen-Hagerling AH2700A Capacitance Bridge in a guarded three-terminal mode at stepped frequencies ranging from 50-20,000 Hz. Two of the terminals were connected to the sample leads shown schematically in Fig. 1a and the third to an electrically isolated copper can surrounding the sample and connected to the ground of the bridge circuit. Most of the measurements were made using 25mV rms excitation, and linearity was confirmed at all fields and temperatures. The bridge was set to output data in the parallel mode in which the sample is assumed to be the circuit equivalent of a capacitance $C'$ in parallel with a resistance $R$ (see Supplemental).

Contacts to the sample were made to the base LPCMO film using pressed indium and to the Al counterelectrode using fine gold wire held in place with silver paint. Silver paint was also occasionally used to make contact to the base electrode with no consequence to the capacitance data at all measurement frequencies. The series combination of the LPCMO parallel resistance with the leakage resistance of the AlOx dielectric was found to be immeasurably large with a lower bound $>10^{10}$ Ω determined by replacing the sample with a standard $10^{10}$ Ω ceramic resistor. Thus the leakage resistance is more than a factor of 1000 higher than the ~10 MΩ maximum resistance of our LPCMO film (Fig. 1b).

The four-terminal resistance measurements of dc transport parallel to the film interfaces (Fig. 1b, black) were made using evenly spaced Van der Pauw contacts (not shown in Fig. 1a) directly connected at the LPCMO film edges. To check for any frequency dependence in the parallel resistance of the LPCMO and associated contact resistance, we performed a two-terminal ac measurement by applying a sinusoidal voltage to all pairs of the LPCMO contacts and used a lock-in amplifier to synchronously detect the output of a wide band current amplifier that provided a return path to ground for the sample current. The two-terminal resistance was quite similar to the four-terminal measurement. Importantly, no frequency dependence in the range 50-20,000 Hz for the resistance varying from 1KΩ to 20MΩ (Fig. 1b) was detected, thus assuring that all the frequency dependence seen in the capacitance measurement is due to perpendicular rather than parallel transport. In addition, we established in separate experiments on symmetric Al/AlOx/Al structures that $C_{AlOx}$ has negligible frequency dispersion over the same frequency range.

**Acknowledgements** We thank Elbio Dagotto, Peter Littlewood and Sergei Obukhov for useful discussions. This work was supported by the National Science Foundation.**Author Information** Correspondence and requests for materials should be addressed to A. F. H. (afh@phys.ufl.edu).Page 14 of 32

**Figure legends:**

**Figure 1 | Parallel (resistance) and perpendicular (capacitance) transport measurements are made on the same LPCMO film. a,** Cross sectional schematic view of the trilayer capacitor structure comprising the LPCMO base electrode, the AlOx dielectric, and the Al counterelectrode. **b,** Semi logarithmic plots of the temperature dependent resistances $R^{\downarrow\uparrow}_{\parallel}(T)$ (black) and capacitances $C^{\downarrow\uparrow}_{\perp}(T)$ (red) for decreasing ($\downarrow$) and increasing ($\uparrow$) temperatures at zero field obtained on the same structure. The thin green lines are fits to the Maxwell-Wagner model (see Supplementary) which incorporates $R^{\downarrow\uparrow}_{\parallel}(T)$ as an input and therefore produces capacitance minima at temperatures coincident with the resistance maxima. At 50 kOe (blue) the capacitance has increased from the zero field minima by a factor of 1000. **c,** Schematic representation of the LPCMO film using distributed circuit elements. Strain effects give rise to a resistance gradient in the perpendicular direction represented schematically by an unequal spacing of equipotential surfaces (dashed horizontal lines) superimposed on a color gradient.

**Figure 2 | Cole-Cole plots reveal data collapse and power-law scaling of the dielectric response. a,** Cole-Cole plot on logarithmic axes of dielectric dissipation $C''(\omega)$ versus polarization $C'(\omega)$ at T = 65 K (warming) for the magnetic fields indicated in the legend. At each field the implicit frequency variable $f$ ranges from 100 Hz to 20 kHz. The crossover from non-overlapping traces at low frequencies to data collapse at high frequencies is magnified in the inset. **b,** Cole-Cole plot for the sweep parameters and

ranges indicated in the legend. Independently of whether the implicit variable is frequency (black squares), temperature (red circles) or magnetic field (blue triangles) the data collapse onto the same curve regardless of the parameter being varied. These same data, replotted as a straight orange line in the inset after subtracting a single fitting parameter $C_\infty$, show power-law scaling collapse (PLSC).

**Figure 3 | Determination of the boundaries of the PLSC region.** Cole-Cole plot on linear axes with temperature (cooling) as the implicit variable for $f = 500$ Hz and magnetic fields identified in the legend. The well-defined transitions onto/(off of) the orange PLSC line (magnified in the inset), which is the same line shown in the inset of Fig. 2b, define respectively the upper/lower field-dependent critical temperatures that bound the PLSC behavior in fTH space shown in Fig. 4.

**Figure 4 | PLSC holds over a wide region of fTH space and at low frequency converges to a point defining the insulator-metal transition $T^{\downarrow}_{IM,\perp}(H=0)$.** The determination of upper (solid symbols) and lower (open symbols) critical temperature lines (cooling, Fig. 3) define triangular-shaped regions in TH space within which PLSC holds for frequencies higher than the frequency (labeled) at which the upper boundary is determined. As $f$ decreases the PLSC region shrinks to a point (large red X) that marks $T^{\downarrow}_{IM,\perp}(H=0)$ and anchors the lower critical temperature line $T^{\downarrow}_{IM,\perp}(H)$ where the capacitance is minimum. The critical temperature line $T^{\downarrow}_{IM,\parallel}(H)$ (green) determined from the peaks in $R^{\downarrow}_{\parallel}(H)$ is roughly parallel to $T^{\downarrow}_{IM,\perp}(H)$ at a temperature offset by ~20K.

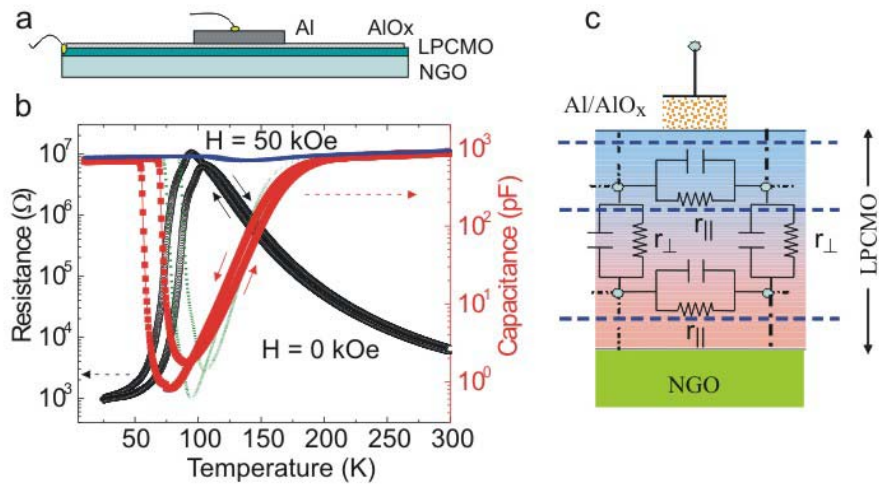

Figure 1

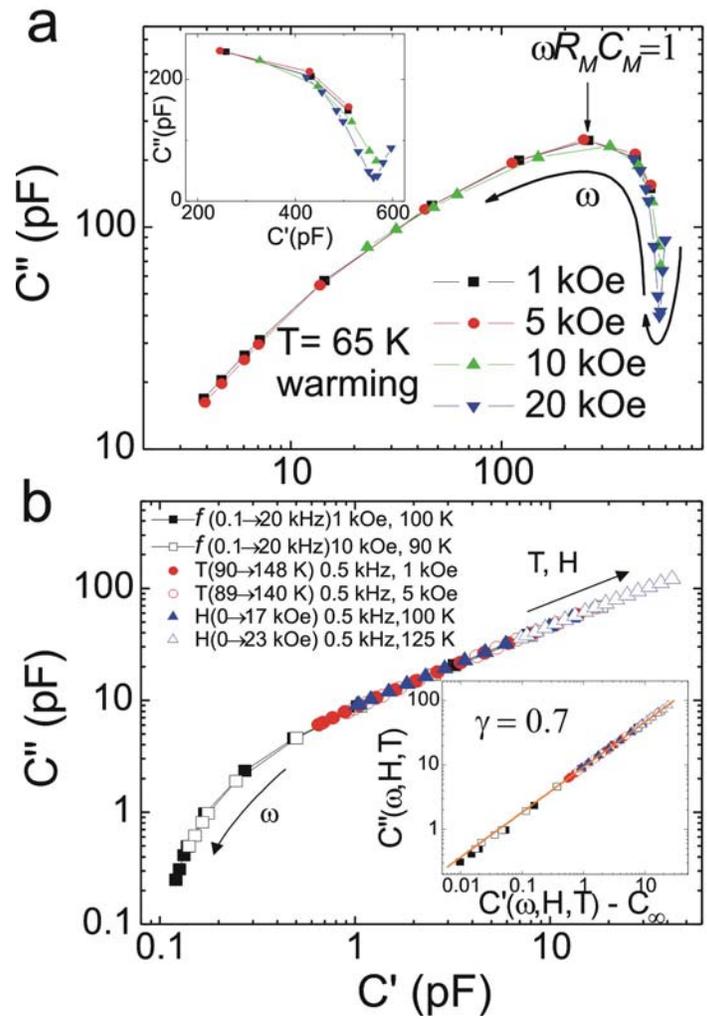

Figure 2

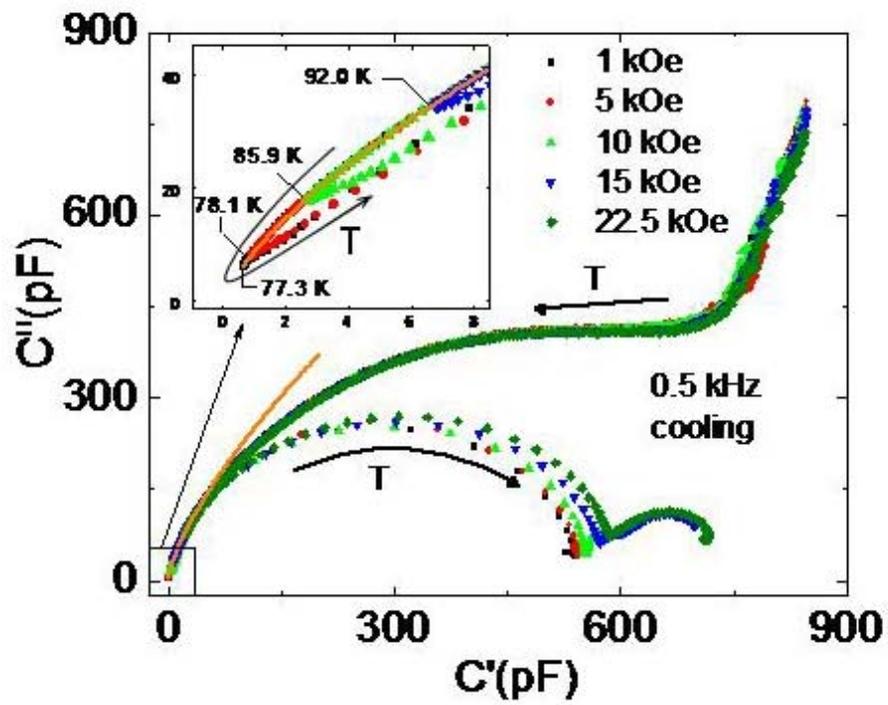

Figure 3

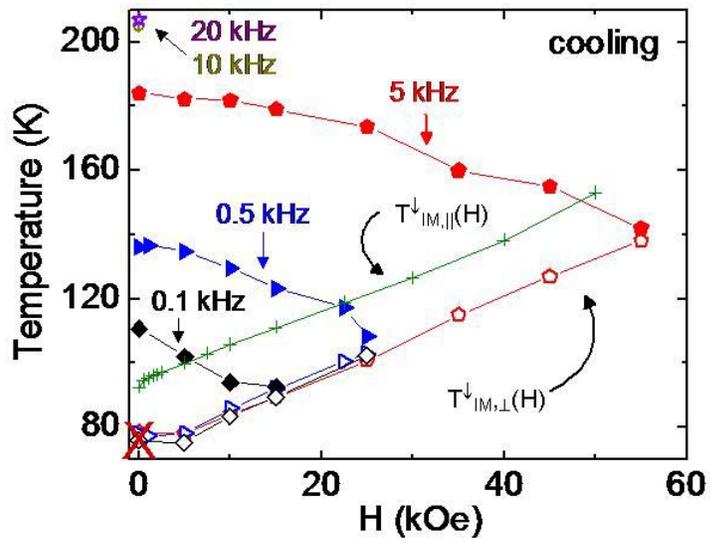

Figure 4

# Supplementary Information

Colossal magnetocapacitance and scale-invariant dielectric response in phase-separated manganites

Ryan P. Rairigh, Guneeta Singh-Bhalla, Sefaatin Tongay, Tara Dhakal, Amlan

Biswas and Arthur F. Hebard

*Department of Physics, University of Florida, Gainesville, FL 32611-8440*

**Overview**

We report on the use of a trilayer configuration in which the sample under investigation, in our case a 600 Å-thick $(La_{1-y}Pr_y)_{5/8}Ca_{3/8}MnO_3$ (LPCMO) film, comprises the base layer of a metal-insulator-metal (MIM) trilayer capacitor structure (see Fig 1a of Letter). Under certain experimental conditions this unconventional configuration allows for the simultaneous measurement of electrical transport both parallel and perpendicular to the film interfaces. Although the four-terminal Van der Pauw measurement of the LPMCO films provides unambiguous information about transport parallel to the film interfaces, the two-terminal capacitance measurement is more problematic, since it includes contributions from both parallel and perpendicular transport. The following Supplementary section augments our Letter by showing that the two-terminal perpendicular contribution to electronic transport can dominate over the parallel contribution providing certain experimental constraints are satisfied. When these conditions are satisfied, we show using the well-known Maxwell-Wagner model that the perpendicular contribution is resolved into two series-connected parts: a contribution from the reference AlOx capacitor and a contribution from the intrinsic dielectric response of the LPCMO film. We then show with additional data on films of different thickness how the substrate strain-induced anisotropy, measured by the difference in temperature between the resistance maxima and capacitance minima, decreases and approaches bulk like behavior as the film thickness increases.

**Comparison of longitudinal and perpendicular voltage drops**

The measured voltage of the two-terminal configuration of Fig. 1a of the Letter can have both parallel and perpendicular contributions from currents flowing respectively either along the LPCMO electrode or transverse to the film through the capacitor. Since these contributions cannot be distinguished in a two-terminal measurement, it is necessary when measuring capacitance to establish conditions where the perpendicular voltage drop dominates over the parallel voltage drop. There are two necessary requirements to assure a dominant perpendicular voltage drop: (1) the dc leakage current through the AlOx dielectric is negligible and (2) the measurement frequency is constrained to be within well defined upper and lower bounds determined by sample properties.

We can begin to understand these requirements by modeling the measurement configuration as a resistance $R_s$ in series with the parallel combination of a complex lossy capacitance, $C^*(\omega) = C_1(\omega) - iC_2(\omega)$, and a dc resistance, $R_0$ (Supplementary Fig. 1a). By lossy capacitance we mean a capacitor that does not pass dc current but does experience loss at ac due to dipole reorientation. Thus the combination of $C^*(\omega)$ shunted by $R_0$ is a leaky capacitor which does pass dc. The resistance $R_s$ includes the parallel resistance $R_\parallel$ of the LPCMO and any resistance associated with the LPCMO contact. The negligible resistance of the Al counterelectrode and its associated contact are included in $R_s$. Our measurements at dc establish the conditions $R_0 + R_s > 10^{10}\,\Omega$ (see Methods section of Letter) and $\max\{R_s\} = 10^7\,\Omega$ (Fig. 1b of Letter), which together imply that over the whole range of dc measurements more than 99.9% of the voltage appears across $C^*$. At temperatures away from the resistance peak this figure of merit improves considerably.

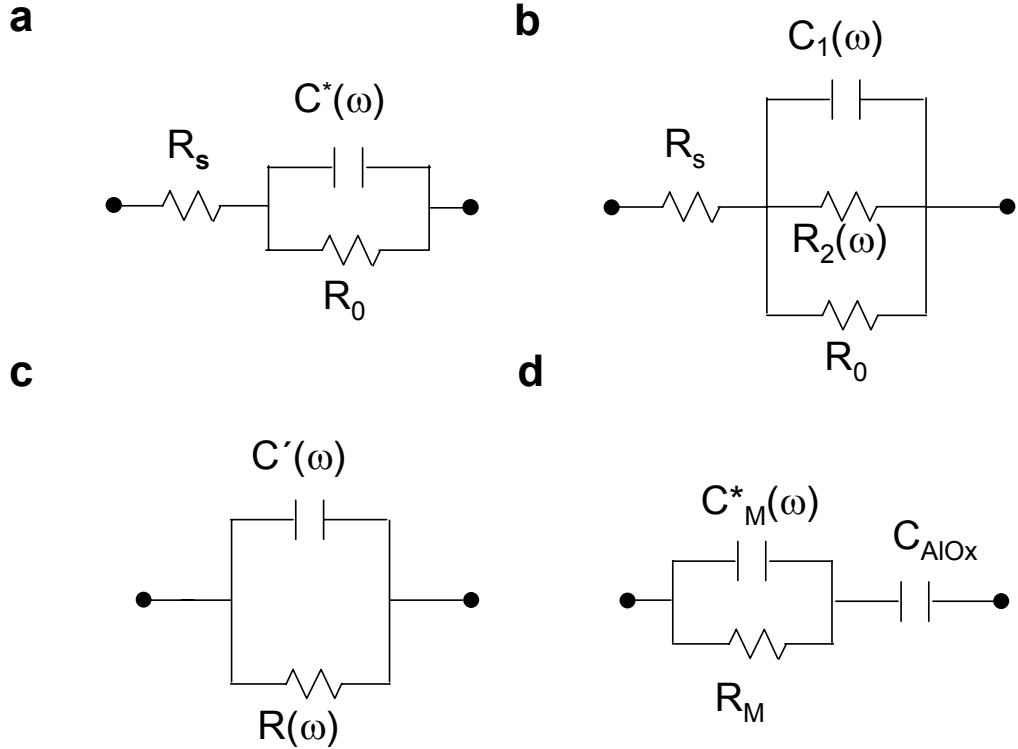

**Supplementary Figure 1 | Circuit diagrams facilitate understanding the sources of longitudinal and perpendicular voltage drops. a,** Circuit equivalent of the two-terminal measurement configuration (Fig. 1a of Letter) where $R_s$ is the series resistance of the LPCMO sample and the parallel combination of a complex (lossy) capacitor $C^*(\omega)$ with a resistor $R_0$ represents the impedance of the LPCMO in series with the aluminum oxide capacitor. In the two-terminal configuration, the longitudinal voltage drop across $R_s$ cannot be distinguished from the perpendicular voltage drop across the parallel combination of $C^*(\omega)$ and $R_0$. **b,** Decomposition of $C^*(\omega) = C_1(\omega) - iC_2(\omega)$ into a parallel combination of $C_1(\omega)$ and $R_2(\omega) = 1/\omega C_2(\omega)$. **c,** Circuit equivalent for the capacitance $C'(\omega)$ and conductance $1/R(\omega)$ reported by the capacitance bridge. **d,** Maxwell-Wagner circuit equivalent for the LPCMO impedance in series with the Al/AlOx capacitor. The LPCMO manganite film impedance is represented as a lossy capacitor $C^*_M(\omega)$ shunted by a resistor $R_M$. There is no shunting resistor across $C_{AlOx}$ because the measured lower bound on $R_0$ is 10 GΩ, well above the highest impedance of the other circuit elements.

Since the capacitance measurements are made at finite frequency, we must consider the more complicated situation of additional current paths and choose conditions to assure that most of the ac potential drop is across $C^*(\omega)$. We do this by redrawing the circuit of Supplementary Fig. 1a to include the ac loss as a resistor $R_2(\omega) = 1/\omega C_2(\omega)$ (Supplementary Fig. 1b) which diverges to infinity at dc ($\omega = 0$). To be sensitive to LPCMO properties, we desire most of the ac current to flow through $R_2(\omega)$ and therefore choose frequencies to satisfy

$$R_2(\omega) = 1/\omega C_2(\omega) \ll R_0 = 10^{10}\,\Omega \;, \tag{S1}$$

thereby determining a lower bound on $\omega$.

The AH capacitance bridge reports the capacitance $C'(\omega)$ and the conductance $1/R(\omega)$ of the parallel equivalent circuit shown in Supplementary Fig. 1c. Using straightforward circuit analysis we relate the measured quantities $C'(\omega)$ and $R(\omega)$ to the circuit parameters of Supplementary Fig. 1b by the equations:

$$C'(\omega) = C_1(\omega)\left(\frac{R_2^2(\omega)}{(R_2(\omega)+R_s)^2 + \omega^2 R_2^2(\omega)R_s^2 C_1^2(\omega)}\right), \tag{S2}$$

and

$$R(\omega) = \frac{(R_2(\omega)+R_s)^2 + \omega^2 R_2^2(\omega)R_s^2 C_1^2(\omega)}{(R_2(\omega)+R_s) + \omega^2 R_2^2(\omega)R_s C_1^2(\omega)} \;. \tag{S3}$$

If $R_s$ is small enough to satisfy the relation

$$R_s \ll \min\left\{\frac{1}{\omega C_1(\omega)}, \frac{1}{\omega C_2(\omega)}, \frac{1}{\omega C_1(\omega)}\left(\frac{C_2(\omega)}{C_1(\omega)}\right)\right\}, \tag{S4}$$

then equations S2 and S3 reduce respectively to $C'(\omega) = C_1(\omega)$ and $R(\omega) = R_2(\omega)$. Accordingly, the fulfillment of the constraints imposed by Eqs. S1 and S4 assures us that the ac dissipation is not due to leakage resistance and that the voltage drop across $R_s$ can be ignored. Under these conditions the measured complex capacitance has real, $C'(\omega)$, and imaginary, $C''(\omega) = 1/\omega R(\omega)$, parts that reflect respectively the polarization and the dissipation plotted and discussed in the Letter.

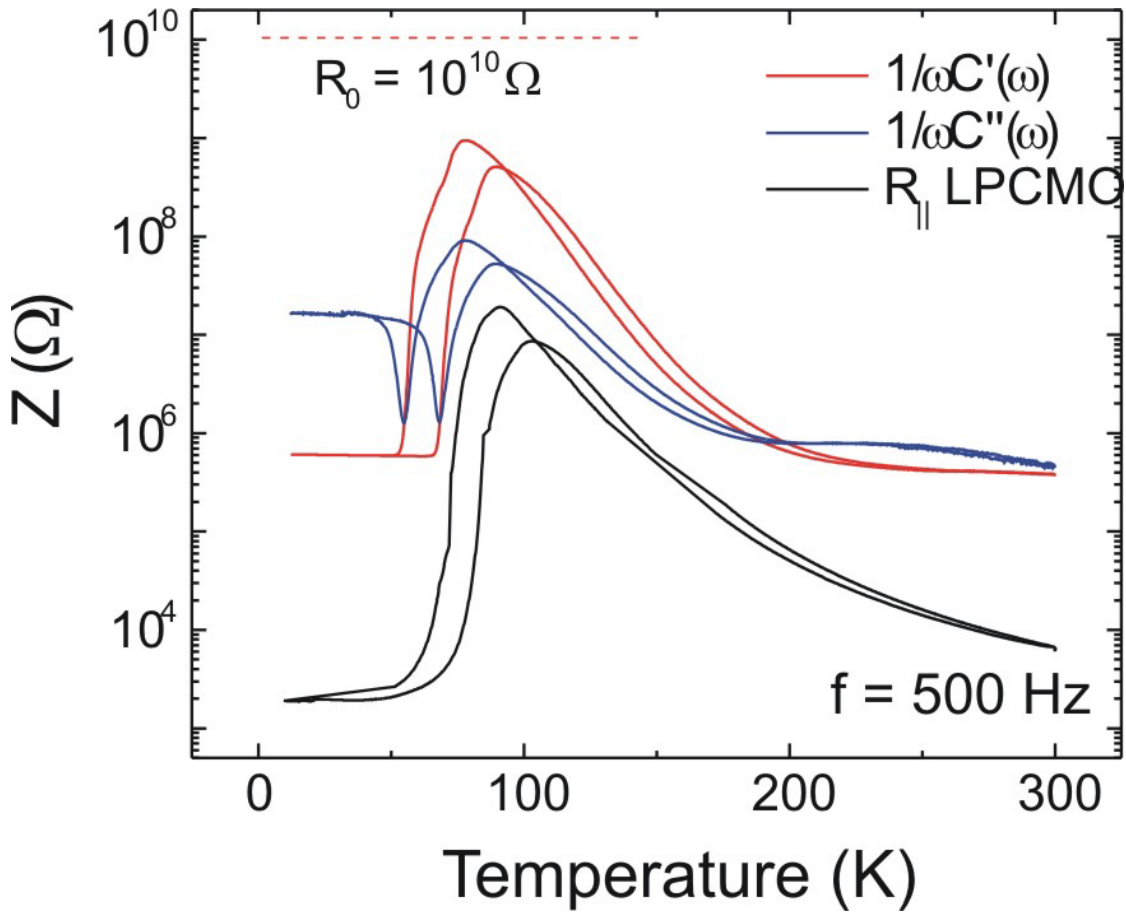

**Supplementary Figure 2 | Impedance plots verify that the longitudinal voltage drops are negligible compared to the perpendicular voltage drops:** The $H = 0$ temperature dependence of the impedance components $1/\omega C'(\omega)$, $1/\omega C''(\omega)$ measured at 500 Hz and $R_s = R_\parallel$ measured at dc. The horizontal dashed line at $10^{10}$ Ω represents the lower bound on $R_0$. Comparison of the relative magnitudes of these plots shows that at all temperatures the constraints imposed by Eqs. S5 and S6 are satisfied.

The constraints of Eqs. S1 and S4 now become

$$1/\omega C''(\omega) \ll R_0 = 10^{10}\,\Omega \tag{S5}$$

$$R_s \ll \min\left\{\frac{1}{\omega C'(\omega)}, \frac{1}{\omega C''(\omega)}, \frac{1}{\omega C'(\omega)}\left(\frac{C''(\omega)}{C'(\omega)}\right)\right\}, \tag{S6}$$

where we have replaced $C_1$ and $C_2$ by the measured quantities $C'$ and $C''$ respectively. These relations conveniently allow us to experimentally determine the range of frequencies over which $R_s$ can be safely ignored, thus guaranteeing that the equipotentials at ac are parallel to the film interface (Fig. 1c of Letter). We show in Supplementary Fig. 2 the $H = 0$ temperature dependence of the impedance components, $1/\omega C'(\omega)$, $1/\omega C''(\omega)$ measured at 500 Hz and $R_\parallel$ measured at dc. The corresponding temperature dependence of $C'(\omega)$ is shown in Fig. 1b of the Letter. Clearly the constraints of (S5) and (S6) are satisfied. It is not necessary to plot the third component of (S6) since $C''(\omega) > C'(\omega)$ for all of our data in the region of collapse (see Figs 2-3 of Letter). We have verified that the constraints hold up to 20 kHz and at all the temperatures and fields used to construct the phase diagram in Fig. 4 of the Letter. For our lowest frequency of measurement (100 Hz) we calculate $C'' = 0.16$ pF as a lower bound below which (S5) cannot be satisfied. For all of our data, $C''(100$ Hz$)$ is more than a factor of ten higher and (S5) is thus satisfied for all of our low frequency data.

**Maxwell-Wagner analysis**

Having established the experimental conditions that allow us to ignore the series voltage drop across $R_s$, we now must distinguish the dielectric responses of the manganite film and the AlOx capacitance. We model $C^*(\omega)$ using a Maxwell-Wagner (MW) circuit equivalent (Ref. 16 of Letter) in which impedance is represented as the series connection of two leaky capacitors. This configuration is often used to account for the effect of contacts in dielectric measurements. In our case the manganite impedance, expressed as a parallel combination of a resistance $R_M$ and capacitance $C_M$, is connected in series with a leak free capacitance $C_{AlOx}$ representing the Al/AlOx circuit element shown in Supplementary Fig. 1d. The resulting expression,

$$C^*(\omega) = C^*_{MW}(\omega) = \frac{C_{AlOx}}{1 + i\omega R_M C_{AlOx}/(1 + i\omega R_M C_M)}, \qquad (S7)$$

reveals a dielectric response determined by two time constants, $R_M C_{AlOx}$ and $R_M C_M$. As ω increases, the capacitance crosses over from being dominated by $C_{AlOx}$ to a capacitance dominated by the series combination of $C_{AlOx}$ and $C_M$, i.e., $C^* = C_M C_{AlOx}/(C_M + C_{AlOx})$. If $C_M \ll C_{AlOx}$, as it is over much of the data range in Fig. 1b of the Letter and likewise for similar data taken in high magnetic fields, then C* in the 'high frequency' limit is equal to $C_M$ and is therefore a direct measure of the LPCMO dielectric response. We test these limits in Fig. 1b of the Letter by evaluating Re$\{C^*_{MW}(\omega)\}$ at 0.5kHz (green curve) using $C_M/C_{AlOx} = 10^{-4}$ and $R_M = R_\parallel(T)^{\downarrow\uparrow}$ (black) as inputs. $C_M$ is assumed to be real for this calculation. The MW model thus provides a good qualitative account of the temperature-dependent capacitance (Fig 1b of Letter, green line) for $C_M$ independent of frequency and equal to $10^{-4} C_{AlOx}$. The MW model also shows good alignment in temperature between the maximum in the resistance used as an input and the calculated capacitance minimum. Finally, we note that the large series-connected aluminum oxide capacitor serves as a reference capacitor, which by its presence 'decloaks' or makes visible the smaller manganite capacitance. If the frequency becomes too high, the constraint (S6) is violated and $R_s$ becomes visible, introducing longitudinal voltage drops that cannot be distinguished from the perpendicular drops.

In reality there is considerable dielectric loss, especially in the presence of magnetic field, and $C_M$ is frequency dependent and therefore complex. If we force $C_M$ in the MW calculation to be complex with, for example, a Debye response, the alignment between the resistance maximum and the capacitance minimum does not change. The Cole-Cole plots (Figs. 2 and 3 of the Letter) are the additional ingredients that clearly capture the interesting intrinsic dynamics of scale invariant dielectric response associated with the interplay of competing phases as discussed in the Letter.

It is worthwhile to further elaborate on intrinsic versus extrinsic effects. The MW model is usually used to ascertain the contributions of contacts and interfaces when the material of interest is sandwiched between two electrodes (Refs 4,5,7,17,18 of Letter). In capacitors with thick dielectrics, the interface region next to either electrode can have distinctly different properties than the interior bulk. Such a heterogeneous system is well described in the MW model by two series-connected leaky capacitors. If one of the leakage components, say the interface, is magnetic field sensitive and exhibits magnetoresistance (MR), then the measured magnetocapacitance (MC) can be a consequence of the extrinsic properties of an interface contact rather than the intrinsic properties of the bulk. In the unconventional configuration described in the Letter, the interface contact is a dispersionless leak-free Al/AlOx capacitor as represented schematically in Fig. S1(d), and the observed MC is due to the intrinsic properties of the mixed phase LPCMO. Any interface effects between the AlOx and the LPCMO are negligible, since the factor of 1000 change in capacitance, which includes the region where power-law scaling collapse is observed, necessarily involves the entire manganite film as described in the concluding section of the Letter. In addition all extrinsic contributions from contacts to the LPCMO at the film edges (Fig 1a of Letter) are included in the resistance $R_s$, which as we have shown above, can be ignored when the frequency is chosen to satisfy the inequality of Eq. S6. Experimentally, this insensitivity was further checked by using silver paint or pressed indium for contacts as described in the Methods section of the Letter.

**Dependence of anisotropy on film thickness**

The insulator to metal (IM) transition in bulk LPCMO is due to a 3D percolation transition. However, as described in the Letter, the presence of strain at substrate/LPCMO interface gives rise to anisotropy as measured by two distinct IM transitions: one in the parallel direction, $T^{\downarrow\uparrow}_{IM,\parallel}$, corresponding to resistance maxima, and the other in the perpendicular direction, $T^{\downarrow\uparrow}_{IM,\perp}$, corresponding to capacitance minima. To verify that the strain-induced anisotropy decreases for thicker more bulk-like films, we have repeated the measurements at zero field for a set of films with three different thicknesses: $d$ = 300Å, 600Å and 900Å.

Figure S3 shows the dependence of $T^{\downarrow\uparrow}_{IM,\parallel}$ and $T^{\downarrow\uparrow}_{IM,\perp}$ on $d$ for cooling and warming as labeled in the legend. For parallel transport the observed increase of transition temperatures can be qualitatively explained by the effect of dimensionality on percolation. Since percolation in 3D occurs at a lower metal fraction than it does in 2D, the IM transition increases with increasing $d$ as is indeed observed. This qualitative picture is complicated however by the presence of a strained layer at the substrate interface which contains a higher fraction of FMM phase. In this case conduction in the parallel direction is facilitated by the presence of the higher conductivity strained layer whereas in the perpendicular direction the current paths must thread regions containing a greater proportion of insulating phase, hence the difference between $T^{\downarrow\uparrow}_{IM,\parallel}$ and $T^{\downarrow\uparrow}_{IM,\perp}$. The temperature differences, $\Delta T^{\downarrow\uparrow}_{IM} = T^{\downarrow\uparrow}_{IM,\parallel} - T^{\downarrow\uparrow}_{IM,\perp}$, for cooling and warming are plotted versus $d$ in Fig. S4. We note that the anisotropy does indeed decrease with increasing $d$. Thus as $d$ increases the IM transition moves to higher temperature and transport becomes more isotropic as the effect of the strained interface diminishes.

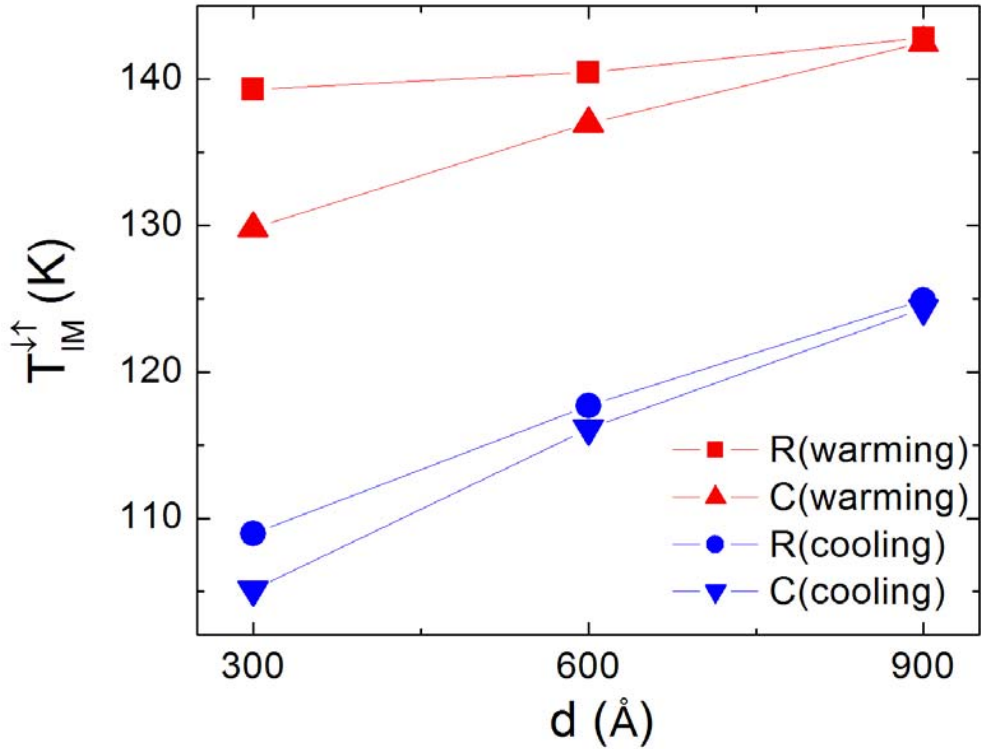

**Supplementary Figure 3 | With increasing LPCMO thickness $d$ the anisotropic IM transitions move to higher temperatures:** The transition temperatures associated with resistance maxima (squares and circles) and capacitance minima (triangles) are identified in the legend and plotted as a function of $d$ for cooling and warming. The capacitance data for the three different films are taken at 100 Hz and satisfy the impedance inequalities expressed in Eqs. S5 and S6 and shown in Fig. S2 for the 600 Å-thick sample described in the Letter.

The films discussed above were prepared from the same target but under different conditions than the 600 Å-thick film discussed in the Letter. The deposition conditions (oxygen pressure = 420 mTorr, substrate temperature = 820 °C, deposition rate = 0.5 Å/s) were determined by minimizing the transition width at an IM transition temperature ($T^{\downarrow}_{IM,\parallel}$) that is close to the maximum value(cooling) observed in bulk compounds of the same composition. The target was then conditioned with the same deposition parameters for many runs. In comparing the two 600 Å-thick films, we see that the transition temperatures $T^{\downarrow}_{IM,\parallel}$ = 117.7K and $T^{\uparrow}_{IM,\parallel}$ = 140.5K of the 'optimized' 600 Å-thick film shown in Fig. S3 are appreciably higher than the corresponding temperatures, $T^{\downarrow}_{IM,\parallel}$=95K and $T^{\uparrow}_{IM,\parallel}$=106K, of the film discussed in the Letter. In addition, the respective anisotropies for cooling ($\Delta T^{\downarrow}_{IM}$ = 1.5K) and warming ($\Delta T^{\uparrow}_{IM}$ = 3.5K) of the 'optimized' 600 Å-thick film are significantly smaller than the corresponding anisotropies for cooling ($\Delta T^{\downarrow}_{IM}$ = 20K) and warming ($\Delta T^{\uparrow}_{IM}$ = 15K) of the same thickness film discussed in the Letter. These results show that our technique can advantageously be used to correlate anisotropies in LPCMO with deposition parameters. We anticipate that this capability will be applicable to other strongly-correlated complex oxide systems as well.

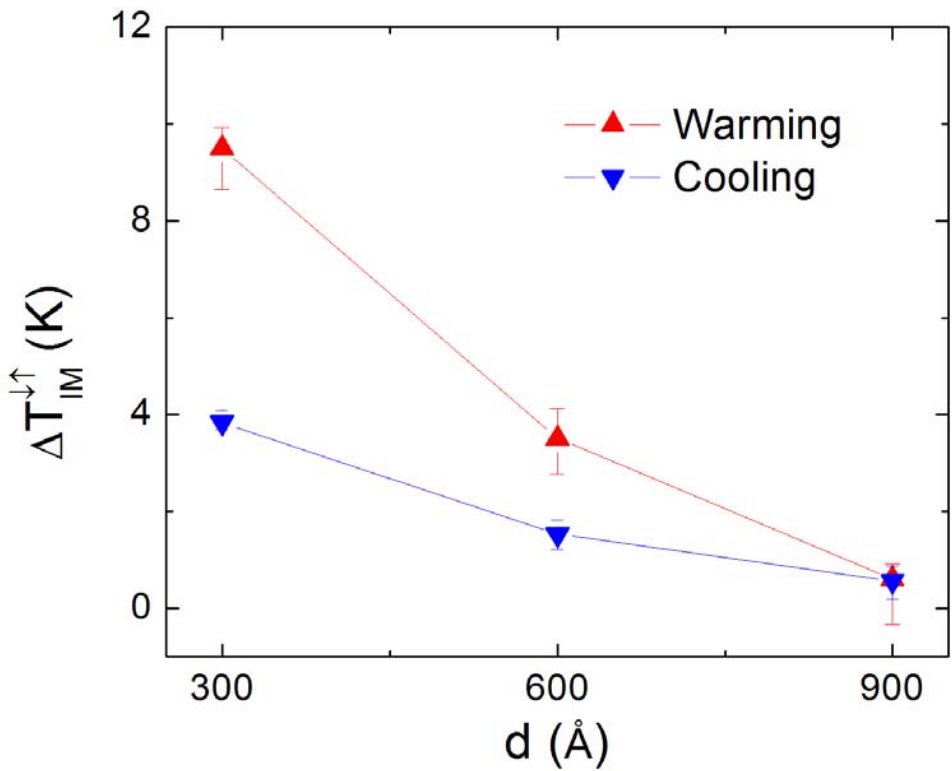

**Supplementary Figure 4 | With increasing LPCMO thickness the anisotropy as measured by $\Delta T^{\downarrow\uparrow}_{IM} = T^{\downarrow\uparrow}_{IM,\parallel} - T^{\downarrow\uparrow}_{IM,\perp}$ decreases towards zero and bulk like behavior:** The data for both cooling and warming cycles at each thickness are obtained from the data shown in Fig. S3 by subtracting the temperature of the capacitance minimum (perpendicular transition) from the temperature of the resistance maximum. The error bars, which are on the order of the symbol size in Fig. S3, are determined by the temperatures which give a ±0.1% deviation at each extremum (resistance maximum or capacitance minimum).